\documentclass[amsmath,amssymb,showpacs,showkeys]{revtex4}
\usepackage{graphicx}
\DeclareMathOperator{\Tr}{Tr}%
\DeclareMathOperator{\e}{e}
\begin{document}
\title{Thermodynamic limit and semi--intensive quantities}
\author{J. Cleymans}
\email{cleymans@qgp.phy.uct.ac.za} \affiliation{UCT-CERN Research Centre and Department of
Physics,University of Cape Town, Rondebosch 7701, South Africa}
\author{K. Redlich}
\email{redlich@ift.uni.wroc.pl} \affiliation{Institute of
Theoretical Physics, University of Wroc{\l}aw, Pl. Maksa Borna 9,
50-204  Wroc{\l}aw, Poland\\
and Gesellschaft f\"ur Schwerionenforschung, D--64291 Darmstadt,
Germany}
\author{L. Turko}
\email{turko@ift.uni.wroc.pl} \affiliation{Institute of Theoretical Physics, University
of Wroc{\l}aw, Pl. Maksa Borna 9, 50-204  Wroc{\l}aw, Poland}
\date{March 16, 2005}
\begin{abstract}
The properties of  statistical ensembles with abelian charges close to the thermodynamic
limit are  discussed. The finite volume corrections to the probability distributions and
particle density moments are calculated. Results are obtained for statistical ensembles
with both  exact and  average charge conservation. A new class of variables
(semi--intensive variables) which  differ in the thermodynamic limit depending on how
charge conservation is implemented in the system is introduced. The thermodynamic limit
behavior of these variables is calculated through the  next to leading order finite
volume corrections to the corresponding probability density distributions.
\end{abstract}
\maketitle
\section{Introduction}\label{intr}

Statistical models have been shown to be very successful in the description of
particle production in heavy ion collisions \cite{brs}. Such  models are usually
constructed from the partition function of  a  quasi non--interacting gas composed
of all known hadrons including hadronic   resonances. The contribution of resonances
is an effective approach to reproduce strong interactions in the hadronic medium
\cite{krh}. The hadron resonance gas model has also been shown to be consistent with
lattice QCD thermodynamics restricted to the confined,  hadronic, phase
\cite{allt,krt,maria}.

Considering a hadronic system using statistical models, it is essential to implement
constraints related to internal symmetries \cite{kt,rh}. We consider here an abelian
symmetry corresponding to one conserved charge. In the statistical system,
conservation laws can be implemented in the canonical (C) or grand canonical (GC)
ensembles. In the following  we consider only ultrarelativistic systems as
encountered e.g. in high energy heavy ion collisions. In such a  case particle
numbers are not conserved, thus there is no associated chemical potentials to the
particle numbers. The only chemical potentials included here are those related to
conserved charges. In the grand canonical ensemble (GC) the quantum numbers have
fixed average values which are determined by the corresponding chemical potentials,
on the other hand, in the canonical ensemble, quantum numbers have fixed values.
This leads to an essential difference in the volume dependence of observables in the
GC and C formulations \cite{kt,rh,rafelski,can,ho}. In the limit when $V\to\infty$
some ratios of extensive quantities converge to different values in the GC and C
ensembles \cite{begun,lt}. The thermodynamic limit (T-limit) is realized if the
volume $V\to \infty$ while the charge density $q$ and the particle densities $n_i$
remain finite in the C system while their thermal average values $\langle q\rangle$
and $\langle n_i\rangle$ in the GC system are kept fixed. It is thus clear that the
equivalence of both descriptions in the thermodynamic limit can be strictly
established only for intensive observables~\cite{brs,rh}. The equivalence of the GC
and C descriptions in the thermodynamic limit has been established at the most
basic, probability density level. It has been shown recently \cite{lt} that
different particle probability densities calculated from the C and GC ensembles
coincide in the thermodynamic limit.

In some physical situations, however, the knowledge of T-limits of probability density
functions is not sufficient. There is a broad class of physical quantities which are
finite in the thermodynamic limit but are still different for different statistical
ensembles. We call henceforth such variables \emph{semi--intensive quantities}. These
variables have finite T-limits, like intensive variables, but those limits are governed
by finite volume corrections to probability distributions for particle densities. The
properties of semi--intensive quantities in the T-limit are entirely determined by the
$\mathcal{O}(V^{-1})$, subleading corrections to the probability density functions. These
contributions are also important if one studies finite volume corrections to
thermodynamic observables. Such a situation appears when comparing the statistical model
with lattice gauge theory results obtained on a small lattice. The statistical model
description of finite volume effects e.g. the charge susceptibilities calculated recently
on the lattice \cite{allt,aeh} require knowledge of the finite volume corrections to the
probability densities.

The phenomenologically  relevant example of a semi--intensive quantity is the scaled
variance
\begin{equation}\label{sc var}
    \omega_N =
\frac{\Delta N^2}{\langle N\rangle} = \frac{\langle N^2\rangle -
\langle N\rangle^2}{\langle N\rangle}\,.
\end{equation}
where $N$ is, e.g., the number of charged particles. The $\omega_N$ is an
experimental observable in heavy ion collisions and  measures the
relative charged particles fluctuation in a system \cite{na49}.

It was recently pointed out \cite{begun} that the scaled variance \eqref{sc var} has
different thermodynamic limits in the GC and C ensembles. In fact, by the substitution
$\langle N^k\rangle=V^k\langle n^k\rangle\,$, the $\omega_N$ is related to the scaled
variance for charged particle densities $\omega_n$ as
\begin{equation}
\omega_N = \frac{\Delta N^2}{\langle N\rangle} =  \frac{V^2\Delta
n^2}{V\langle n\rangle}\equiv V\omega_n\,.
\end{equation}
In the  T-limit the scaled variance $\omega_n$ vanishes both in the GC and in the C
ensembles. We show that this can be formally expressed as
\begin{equation}
    \omega_n = \frac{1}{V}\mathcal{R}+\mathcal{O}(V^{-2})\,.
\end{equation}
Consequently the T-limit of  the scaled variance $\omega_N$ is sensitive to the next to
leading order (NLO) corrections $\mathcal{O}(V^{-1})$. The subleading term $ \mathcal{R}$
is specific to a given ensemble, that is why the T-limit of $\omega_N$ is ensemble
dependent.

In this paper we calculate the NLO corrections to different probability distribution
functions in the C and the GC ensembles. These subleading corrections will be explicitly
determined in the hadron resonance gas model constrained by  charge conservation. We
will show that in the T-limit the probability density distributions do not exist as
regular but appear as generalized functions. We also calculate the T-limit properties of
the related particle density moments. Finally, as one of the applications, we introduce a
class of semi--intensive variables in the canonical and the grand canonical ensemble and
discuss their behavior in the vicinity to the thermodynamic limit.

\section{Distributions and moments in finite systems}

The properties of charged particle  probability distributions  in their approach to the
thermodynamic limit will be discussed in the context of the statistical model of a
non--interacting  gas constrained by the conservation of the abelian charge $Q$. The
thermodynamic system of volume $V$ and temperature $T$ is considered to be composed of
charged particles and their antiparticles carrying charge $\pm 1$ respectively. The
requirement of charge conservation in the system is imposed on the grand canonical  or
canonical  level.

The partition function of  the above  C and GC statistical system is  found to be
\begin{subequations}
  \begin{align}\label{part funct}
\mathcal{Z}_Q^{C}(V,T)&=\Tr_Q\,\e^{-\beta\hat{H}}= I_Q(2
z)\,,\\
\mathcal{Z}^{GC}(V,T)&=\Tr\e^{-\beta(\hat{H} -\mu
\hat{Q})}=\exp{\left(2z\cosh{\frac{\mu}{T}}\right)}\,.
\end{align}
\end{subequations}
where $z$ is the sum over all  one-particle partition functions
\begin{equation}
z(T)=\frac{V}{(2\pi)^3}\sum_i g_i\int d^3p\,\e^{-\beta\sqrt{p^2 +m_i^2}}=
\frac{V}{2\pi^2}T\sum_i g_i\,m_i^2\,K_2\left(\frac{m_i}{T}\right)\equiv V z_0(T)\,,
\end{equation}
and $g_i$ is the spin degeneracy factor. The sum is taken over all charged particles and
resonances of mass $m_i$ carrying the charge $\pm 1$. The functions  $I_Q$ and $K_2$ are
modified Bessel functions. The chemical potential $\mu$ determines the average charge in
the GC ensemble.

Semi--intensive variables are constructed from different particle moments $\langle
N^k\rangle$ and their  volume dependence is obtained from the corresponding behavior of
the particle number probability distribution.

In the C ensemble of a system of volume $V$ and  total charge $Q$ the probability
distribution $\mathcal{P}_Q^{C}(N,V)$ to have  $N$ negatively  and $N+Q$ positively
charged particles is obtained \cite{lt,kkl} from the partition function \eqref{part
funct} as
 \begin{equation}\label{prob C N}
    \mathcal{P}_Q^{C}(N,V)= \frac{z^{2N+Q}}{N!(N+Q)!}\frac{1}{I_Q(2 z)}\,.
\end{equation}
On the other hand  in the GC ensemble with volume $V$ and  average charge
$\langle Q\rangle $ the probability distribution
$\mathcal{P}_{\langle Q\rangle}^{GC}(N,Q,V)$ to
find a system with a given charge $Q$ and a given number of negatively charged particles
$N$ is expressed \cite{lt} as the product
\begin{equation}\label{prob GC NQ}
\mathcal{P}_{\langle Q\rangle}^{GC}(N,Q,V) =
\mathcal{P}_Q^{C}(N,V)\,\mathcal{P}_{\langle
Q\rangle}^{GC}(Q,V)\,,
\end{equation}
of the canonical particle number distribution
$\mathcal{P}_Q^{C}(N,V)$ from Eq. \eqref{prob C N} and the grand
canonical probability distribution
\begin{equation}\label{prob GC Q}
    \mathcal{P}_{\langle Q\rangle}^{GC}(Q,V)=
  I_Q(2z)\left[\frac{\langle Q\rangle +
  \sqrt{\langle Q\rangle^2+4 z^2}}{2 z}\right]^Q \e^{-\sqrt{\langle Q\rangle^2+4  z^2}}\,
\end{equation}
to find the total charge $Q$ in the system with the average charge
$\langle Q\rangle$.

 With the knowledge of the probability distributions from Eqs. \eqref{prob C N} and \eqref{prob GC Q}
 the thermal average of particle moments in the GC and C ensemble  are  obtained from
\begin{subequations}\label{mom GC k}
  \begin{eqnarray}
\langle N^k\rangle^{GC}_{\langle Q\rangle}&=&\sum_{N=0}^\infty\sum_{Q=-N}^\infty
N^k\mathcal{P}_{\langle Q\rangle}^{GC}(N,Q,V)\,,\\
\langle N^k\rangle^{C}_{ Q}&=&\sum_{N=0}^\infty N^k\mathcal{P}_{ Q}^{C}(N,V)\,.
\end{eqnarray}
\end{subequations}
In the following section we will discuss the generalizations of the above results for the
probability distributions and particle moments  that are  required to analyze the
approach to the thermodynamic limit.

\section{Distributions  and moments in infinite systems}

In the previous section we have summarized  how to relate particle moments with
probability distributions in a system that is constrained by charge conservation. These
results, however, are only valid for finite systems far away from the thermodynamic
limit. Obviously in the T-limit different particle moments introduced in Eq. \eqref{mom
GC k},  as well as the particle number,  the total charge  and  their average values
appearing in Eqs. \eqref{part funct} and \eqref{mom GC k} are  all infinite. Thus, to
take  the thermodynamic limit in Eqs. \eqref{part funct} and \eqref{mom GC k} one first
expresses the variables $(N,Q,N^k)$ by means of the corresponding densities $(n,q,n^k)$
and then takes the limit $V\to \infty$ keeping the densities fixed. This also requires
the replacement of the discrete sums $(1/V)\sum_N \to \int dn$ and $(1/V)\sum_Q \to \int
d\rho$ by the corresponding integrals over densities.

To formulate correctly  the thermodynamic limit of quantities involving
densities, one defines  the following
probabilities
\begin{subequations}
  \begin{eqnarray}
{\mathbf{P}}_{q}^{C}(n,V) &:= & V\mathcal{P}_{V q}^{C}(V
n,V)\,,\\
{\mathbf{P}}_{\langle  q\rangle}^{GC}( n,q,V)&:= &
V^2\mathcal{P}_{V\langle q\rangle}^{GC}(V n,V q,V)\,,\\
{\mathbf{P}}_{\langle  q\rangle}^{GC}(q,V)&:= &
V\mathcal{P}_{V\langle q\rangle}^{GC}(V q,,V)\,.\label{prob GC
dens}
 \end{eqnarray}
\end{subequations}
such that in the limit $V\to \infty$
\begin{subequations}\label{prob corrts}
  \begin{eqnarray}
    {\mathbf{P}}^{C}_q( n,V) &=&
    \mathcal{P}_{q}^{\infty}(n)+\frac{1}{V}R^{C}_q(n)+\mathcal{O}(V^{-2})\,,\label{prob corrts C}\\
    {\mathbf{P}}^{GC}_{\langle  q\rangle}(n,q,V) &=&
    \mathcal{P}_{\langle  q\rangle}^{\infty}(n,q)+\frac{1}{V}R^{GC}_{\langle q\rangle}(n,q) +
    \mathcal{O}(V^{-2})\,,\label{prob corrts GC nq}\\
    {\mathbf{P}}^{GC}_{\langle  q\rangle}(q,V) &=&
    \mathcal{P}_{\langle  q\rangle}^{\infty}(q)+\frac{1}{V}S^{GC}_{\langle q\rangle}(q) +
    \mathcal{O}(V^{-2})\,.\label{prob corrts GC q}
\end{eqnarray}
\end{subequations}
The first terms    $\mathcal{P}_{q}^{\infty}(n)$,    $
\mathcal{P}_{\langle q\rangle}^{\infty}(n)$ and
${\mathcal{P}}^{\infty}_{\langle q\rangle}(q) $ are the T-limits
distributions corresponding to the $V\to \infty$ limit, the second
terms in Eq. \eqref{prob corrts}  are the finite volume NLO
corrections. Obviously,  the relation \eqref{prob GC NQ} between
GC and C probabilities also holds in the vicinity of the
thermodynamic limit. Thus, the probability distribution
\eqref{prob corrts GC nq} is just a product of C and GC
probabilities from Eqs. \eqref{prob corrts C} and \eqref{prob
corrts GC q}.

With the above parametrization of the probability distributions,  the particle density
moments in the C and the GC ensembles are obtained from
\begin{subequations}\label{mom dens}
  \begin{align}
    \langle n^k\rangle_{q}^C& =
    \int dn\,n^k\,\mathcal{P}_{q}^{\infty}(n)+ \frac{1}{V}\int dn\,n^k\,R^{C}_q(n)+
    \mathcal{O}(V^{-2})\,,\label{mom dens C}\\
    \langle n^k\rangle_{\langle q\rangle}^{GC}& =
    \int dn\,n^k\int dq\,\mathcal{P}_{\langle  q\rangle}^{\infty}(n,q)+
    \frac{1}{V}\int dn\,n^k\int dq\,R^{GC}_{\langle q\rangle}(n,q) +
    \mathcal{O}(V^{-2})\label{mom dens GC}\,.
\end{align}
\end{subequations}
The equivalence of the GC and C ensembles in the thermodynamic limit requires that the
first terms in the above equations coincide. Indeed it was shown \cite{lt} that  the
charged density distribution function ${\mathbf{P}}_{\langle q\rangle}^{GC}(q,V)$
converges to the Dirac delta function such that
\begin{equation}\label{prob q limit}
 \mathcal{P}_{\langle  q\rangle}^{\infty}(n,q)=
 \mathcal{P}_{q}^{\infty}(n)\cdot\delta (q-\langle q\rangle).
\end{equation}
 The above relation establishes the equivalence of  the GC and C ensembles in
the thermodynamic limit on the most general, probability level. Indeed, substituting Eq.
\eqref{prob q limit} into Eq. \eqref{mom dens} it is clear that any  charged particle
density moment converges to the same asymptotic value $\langle n^k\rangle_\infty$ in the
GC and C ensemble. In the T-limit the charge density $q$ is also identified with its
thermal average value $\langle q\rangle$ following  Eq. \eqref{prob q limit}.

The coincidence  of  C and  GC probability distributions and the corresponding charged
particle density moments in the asymptotic limit of  $V\to \infty$ is not any more valid
for  the NLO contributions. The correction coefficients of order $(1/V)$ in Eq.
\eqref{prob corrts} and \eqref{mom dens} are in general different in the C and GC
ensembles.

For physical applications it is important to know the NLO behavior of the probability
functions. The asymptotic properties of e.g. the semi--intensive  quantities  will be
shown in the next section to be  sensitive to the finite volume corrections appearing in
Eqs. (\ref{prob corrts}--\ref{mom dens}).

In the following we will discuss how to obtain the NLO contributions to the probability
distributions. We will then apply these results to establish the properties of particle
density moments as well as semi--intensive quantities in their approach towards the
thermodynamic limit.
\subsection{Finite volume corrections to the probability distributions and density moments  in the C ensemble }

To obtain the finite volume corrections to the probability
distribution \eqref{prob corrts} and the corresponding charged
particle moments \eqref{mom dens} it is convenient to introduce a
generating function
\begin{equation}\label{gen fn}
    \mathcal{G}(\lambda,V)=\sum\limits_{N=0}^\infty \lambda^N {\mathcal{P}}^{C}_Q( N,V)\,.
   \end{equation}
The density moments $ \langle N^k\rangle^C$ are obtained from the generating function
\eqref{gen fn} as
  \begin{equation}\label{gen fn mom}
        \langle
    N^k\rangle^C=\left.\left(\lambda\frac{d}{d\lambda}\right)^k \mathcal{G}(\lambda,V)\right|_{\lambda=1}\,.
\end{equation}
For the probability distribution  in the canonical ensemble \eqref{prob C N}
 the generating function  has the following form
\begin{equation}\label{gen fn C}
    \mathcal{G}(\lambda,V)
    =\lambda^{-\frac{Vq}{2}}\,
    \frac{I_{Vq}(2V z_0\sqrt\lambda)}{I_{ Vq}(2Vz_0)}\,.
\end{equation}
The T-limit behavior of particle moments and the probability distribution is  obtained
from Eqs. \eqref{gen fn mom} and \eqref{gen fn C} using the  asymptotic behavior of the
generating function for  $V\to \infty$ with  fixed $q$. This is determined by the
limiting properties of the Bessel function \cite{Abram:2004zb}
\begin{equation}\label{bess as}
\lim_{\alpha \to \infty} I_\alpha(\alpha z)=
    \frac{\e^{\alpha\sqrt{1+z^2}}}{\sqrt{2\pi\alpha}(1+z^2)^{1/4}}
    \left[\frac{z}{1+\sqrt{1+z^2}}\right]^\alpha\left\{1+\frac{3z^2-2}{24(1+z^2)^{3/2}\alpha}
    +\mathcal{O}(\alpha^{-2})\right\}.
\end{equation}
The generating function \eqref{gen fn C} in the T-limit is now
obtained as
\begin{equation}\label{gen fn as}
    \mathcal{G}(\lambda,V)=
   \sqrt{ \frac{x_q}
    {x_{q;\lambda}}}\e^{V(x_{q;\lambda}-x_q)}
    \left[\frac{ q+x_q}
    { q + x_{q;\lambda}}\right]^{V q}\times
    \left\{1+\frac{6 z_0^2\lambda -
      q^2}{12 Vx_{q;\lambda}^{3}} - \frac{6 z_0^2 -
      q^2}{12 V x_q^{3}}+\mathcal{O}(V^{-2})\right\}\,.
\end{equation}
where we have introduced
\begin{equation}\label{coeff x}
    x_q=\sqrt{q^2+4z_0^2}\,,\qquad x_{q;\lambda}=\sqrt{q^2+4
z_0^2\lambda}\,.
\end{equation}
Applying  the above  asymptotic form  of the generating function
in  Eq. \eqref{gen fn mom} one establishes (for details see
Appendix \ref{C asympt}) the T-limit  of  positively
and negatively charged particle density moments in the canonical
ensemble
\begin{equation}\label{mom dens C lim}
      \langle n_\pm^k\rangle^C=\langle n_\pm\rangle_\infty^k
      -\frac{k}{V}\frac{z_0^2}{x_q^2}\langle n_\pm\rangle_\infty^{k-1}+
    \frac{k(k-1)}{2V}\frac{z_0^2}{x_q}\langle n_\pm\rangle_\infty^{k-2}
    +\mathcal{O}(V^{-2})\,,
\end{equation}
with $x_q$ defined  as in Eq. \eqref{coeff x}. The average charged particle densities
\begin{equation}\label{dens}
\langle n_\pm\rangle_\infty=\frac{\sqrt{q^2+4z_0^2}\pm q}{2}
\end{equation}
are  the asymptotic results obtained from the $V\to \infty $ limit and have common values
in the C and the GC ensembles if one identifies, following Eq. \eqref{prob q limit}, the
charge density $q$ with its thermal average value $\langle q\rangle$.

Taking into account   Eq. \eqref{mom dens C} together with  the $(1/ V)$ expansion
of the charged particle moments \eqref{mom dens C lim} one finds that  the
probability density ${\mathbf{P}}^{C}_q(n;V)$ of negatively charged particles in the
near vicinity to the thermodynamic limit reads
\begin{equation}\label{eaq22}
      {\mathbf{P}}^{C}_q(n;V)=
      \delta\left(n-\langle n\rangle_{\infty}\right)+
      \frac{1}{V}\frac{z_0^2}{x_q^2}\,\delta'\left(n-\langle n\rangle_{\infty}\right)+
      \frac{1}{V}\frac{z_0^2}{2x_q}\,\delta''\left(n-\langle n\rangle_{\infty}\right)+
      \mathcal{O}(V^{-2})\,,
\end{equation}
where the prime and double prime in the delta functions denote the
first and the second order derivatives with respect to the particle
density $n$.

Eqs. \eqref{mom dens C lim} and \eqref{eaq22} summarize the T-limit behavior of the
charged particle density distribution and the density  moments in the canonical
system with exact charge conservation.

\subsection{Finite volume corrections  to   the probability distributions and density moments  in the GC ensemble }
To establish the asymptotic behavior of the GC probability distribution of charged
particle density and the density moments in the near vicinity to thermodynamic limit
we use the following decomposition of the probability density function
${\mathbf{P}}^{GC}_{\langle q\rangle}( q,n;V)$ \cite{lt}
\begin{equation}
            {\mathbf{P}}^{GC}_{\langle q\rangle}( q,n;V) =
{\mathbf{P}}^{GC}_{\langle q\rangle}( q;V){\mathbf{P}}^{C}_q(
n;V)\,.
\end{equation}
The T-limit is obtained from the corresponding behavior of
${\mathbf{P}}^{GC}_{\langle q\rangle}( q;V)$ and
${\mathbf{P}}^{C}_q( n;V)$ distributions. From Eq. \eqref{prob
corrts} one gets
\begin{equation}\label{eaq24}
  {\mathbf{P}}^{GC}_{\langle  q\rangle}( q,n;V) = \mathcal{P}_{\langle  q\rangle}^{\infty}(q)
  \mathcal{P}_{q}^{\infty}(n)\,+  \frac{1}{V}\left(S^{GC}_{\langle
q\rangle}(q)\mathcal{P}_{q}^{\infty}(n)+R^{C}_q(n)\mathcal{P}_{\langle
q\rangle}^{\infty}(q)\right)+\mathcal{O}(V^{-2})\,.
\end{equation}
The  coefficient in the  $(1/V)$--expansion   of the canonical probability density
${\mathbf{P}}^{C}_q( n;V)$ up to $\mathcal{O}(V^{-2})$ is  already known from Eq.
\eqref{eaq22}. However, the GC probability distribution ${\mathbf{P}}^{GC}_{\langle
q\rangle}( q;V)$ is  only known in the leading order \cite{lt}
\begin{equation}
{\mathbf{P}}^{GC}_{\langle q\rangle}( q;V)=  \delta(q-\langle
 q\rangle)+O(1/V)\,.
\end{equation}
The explicit derivation of the
$(1/V)$--corrections to ${\mathbf{P}}^{GC}_{\langle q\rangle}$
is given in Appendix B where it is  shown that
\begin{equation}\label{eaq26}
{\mathbf{P}}^{GC}_{\langle  q\rangle}(q,V)=
  \delta\left( q-\langle q\rangle\right) +
  \frac{x_{\langle q\rangle}}{2V  }\,\delta^{''}( q -\langle q\rangle)
  +\mathcal{O}(V^{-2})\,,
\end{equation}
with $x_{\langle q\rangle}$ as in Eq. \eqref{coeff x} but with
$q$ being replaced by $\langle q\rangle$.

Substituting the asymptotic expansions  \eqref{eaq22} and \eqref{eaq26} to Eq.
\eqref{eaq24} one gets after some functional algebra the grand canonical probability
density as
\begin{equation}
{\mathbf{P}}^{GC}_{\langle  q\rangle}( q,n;V)=
    \delta\left(n-\langle n\rangle_{\infty}\right)\delta(q-\langle q\rangle) +
    \frac{\langle n\rangle_{\infty}}{2V}
    \delta''\left(n-\langle n\rangle_{\infty}\right)\delta(q-\langle q\rangle)+\mathcal{O}(V^{-2})\,.
\end{equation}
where $\langle n\rangle_{\infty}$ is the average density
in the  $V\to \infty $ limit which is common for both ensembles.
Carrying out the charge integration in the above equation one gets the GC particle number
density probability distribution

\begin{equation}\label{eaq28}
{\mathbf{P}}^{GC}_{\langle  q\rangle}( n,V)=
    \delta\left(n-\langle n\rangle_{\infty}\right) +
    \frac{\langle n\rangle_{\infty}}{2V}
    \delta''\left(n-\langle n\rangle_{\infty}\right)+\mathcal{O}(V^{-2})\,.
\end{equation}
The above result can be directly compared with the corresponding probability
distribution \eqref{eaq22}  in the C ensemble. It is clear that the leading terms in
both ensembles coincide. This is to be expected due to equivalence of the C and GC
ensembles in this limit. However, the $(1/V)$--corrections in both ensembles are
obviously different. This indicates that the C and GC probability distributions
converge to the asymptotic, $V\to \infty $ limit with a different strength. That is
why,  in any finite volume the thermodynamic observables  calculated  in both these
ensembles have, in general, different values. %

\begin{figure}[h]
  \includegraphics[width=5.9cm]{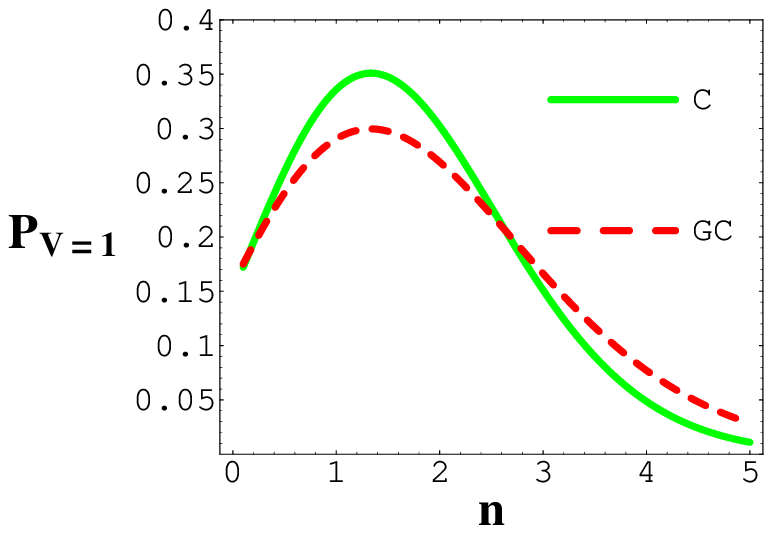}
  \includegraphics[width=5.9cm]{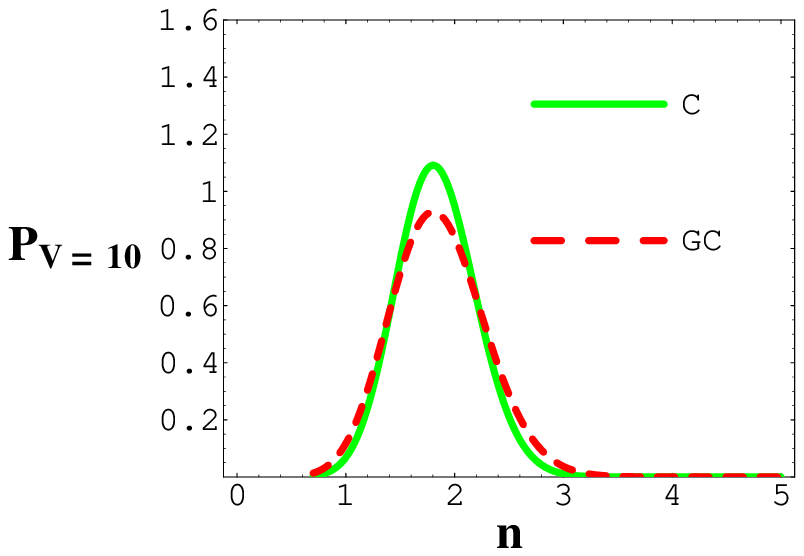}
  \includegraphics[width=5.9cm]{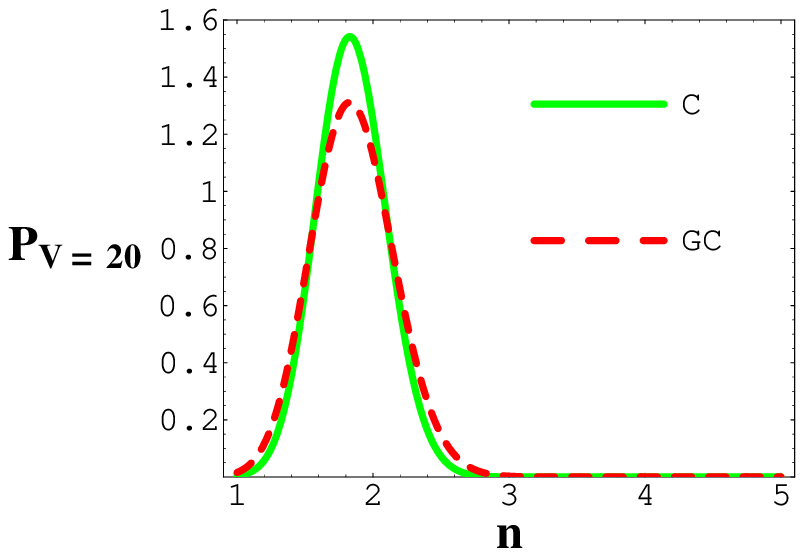}
  \caption{Probability distributions of charged particle density in the canonical
and grand canonical ensembles  calculated for different  volume parameters (V=1, 10
and 20 in arbitrary units).} \label{distrib}
\end{figure}

The properties of negatively charged particle number density distributions
${\mathbf{P}}^{C}_q(n;V)$ and ${\mathbf{P}}^{GC}_{\langle q\rangle}(n,V)$ calculated
by means of exact formulae \eqref{prob C N}--\eqref{prob GC Q} are illustrated in
Fig.~\ref{distrib} for different volume parameters. It is seen in Fig.~\ref{distrib}
that the largest differences between C and GC distributions appear for small
volumes. With increasing $V$ both distributions become narrower, however
${\mathbf{P}}^{GC}_{\langle q\rangle}$  is always broader than ${\mathbf{P}}^{C}_q$.
In the large $V\to\infty$ limit both these distributions converge to  generalized
functions which are explicitly described through Eqs. \eqref{eaq22} and
\eqref{eaq28}.

Knowing the asymptotic properties of  the probability distributions on can study the
corresponding  behavior of the positively and negatively charged particle density
moments. The asymptotic behavior of the positively and negatively charged particle
density moments in the GC ensemble can be obtained from Eq. \eqref{mom dens GC} and
\eqref{eaq28} in the following transparent form

\begin{equation}\label{eaq29}
      _\langle {n_\pm^k}\rangle^{GC}=\langle n_\pm\rangle_{\infty}^k +
      \frac{k(k-1)}{2V}
      \langle n_\pm\rangle_{\infty}^{k-1}+\mathcal{O}(V^{-2}),
\end{equation}
where $\langle n_\pm\rangle_{\infty}$ is as before the average density of $(\pm 1)
$-- charged particles in the  $V\to \infty $ limit which is common for   GC and   C
ensemble. In a particular case, for  the second moment $(k=2)$,  the
Eq.\eqref{eaq29} multiplied by $V^2$ is  a well known result from statistical
mechanics:  $\langle N^2\rangle -\langle N\rangle^2= \langle N\rangle$ which tells
us  that  the average square fluctuations in particle number in the GC ensemble are
controlled by the average number of particles.

\section{Semi-intensive quantities in the thermodynamic limit }

As an application of the results for the asymptotic behavior of the probability
distributions and charged particle moments we consider the properties of semi--intensive
quantities. The simplest example of such a quantity, as discussed in the introduction, is
the  scaled variance of  charged particles

\begin{equation}\label{eaq30}
    \omega_{N_\pm} =
  \frac{\langle N_\pm^2\rangle -
\langle N_\pm\rangle^2}{\langle N_\pm\rangle}\,.
\end{equation}
Obviously, one can express $ \omega_{N_\pm}$ through the corresponding density moments
$\omega_{n_\pm}$  such as
\begin{equation}\label{eaq31}
    \omega_{N_\pm} =
  V\frac{\langle n_\pm^2\rangle -
\langle n_\pm\rangle^2}{\langle n_\pm\rangle}\equiv V\omega_{n_\pm}\,.
\end{equation}
The thermodynamic limit of $ \omega_N$ is controlled by the $(1/V)$-- corrections to
the first and second moment of the charged particle density. From the results in the
previous section it is clear that in the T-limit
\begin{equation}\label{eaq32}
    \langle n^k\rangle=\langle n\rangle_{\infty}^k +
   \frac{1}{V}\mathcal{R}_{k;q}+\mathcal{O}(V^{-2})\,,
\end{equation}
where the first term is common for the GC and C ensembles, however the $(1/V)$
correction term $\mathcal{R}_{k;q}$   differs in these ensembles. Indeed, from Eq.
\eqref{mom dens C lim} and \eqref{eaq29} one finds
\begin{eqnarray}
  \mathcal{R}_{k;q}^C &=&
\frac{k(k-1)}{2}\frac{z_0^2}{x_q}\langle n\rangle_{\infty}^{k-2}
  -k\frac{z_0^2}{x_q^2}\langle n\rangle_{\infty}^{k-1}\label{eaq33}
 \,,\\
  \mathcal{R}_{k;q}^{GC} &=& \frac{k(k-1)}{2}\langle
  n\rangle_{\infty}^{k-1}\label{eaq34}\,,
\end{eqnarray}
where $\mathcal{R}^C$ and  $\mathcal{R}^{GC}$ are the correction terms in  C and GC
ensemble respectively.

From Eq. (\ref{eaq32}) it is clear that the ratio

\begin{equation}\label{estim}
    \Delta_k=\frac{\mathcal{R}_{k;q}}{V\langle n\rangle_{\infty}^k}\,,
\end{equation}
describes the NLO corrections to the $k$-th order density moment. Following Eqs.
(\ref{eaq33}) and (\ref{eaq34}) one finds that these corrections for the canonical
ensemble

\begin{equation}\label{new1}
    \Delta^C_k = \frac{k(k-1)}{V\langle n\rangle_{\infty}}\,\frac{z_0^2}{2 x_q\langle n\rangle_{\infty}}
  -\frac{k}{V\langle n\rangle_{\infty}}\,\frac{z_0^2}{x_q^2}\,,
\end{equation}
and for the grand canonical ensemble
\begin{equation} \label{new2}
    \Delta^{GC}_k = \frac{k(k-1)}{V\langle n\rangle_{\infty}}\,,
\end{equation}
are expressed through the total number of particles of a given charge ${V\langle
n\rangle_{\infty}}$  as well as through dimensionless variables $z_0^2/x_q\langle
n\rangle_{\infty}$ and $z_0^2/x_q^2$ which depend only on the $q/z_0$ ratio.

From  Eq. (\ref{dens}) one finds
\begin{equation}\label{var ratio}
    \frac{q^2}{z_0^2} = \frac{\langle n_+\rangle_{\infty}}{\langle
    n_-\rangle_{\infty}} + \frac{\langle n_-\rangle_{\infty}}{\langle
    n_+\rangle_{\infty}} - 2\,,
\end{equation}
thus the NLO corrections $\Delta_k$ can be  expressed by means of
measurable $\langle N_-\rangle/\langle N_+\rangle$  particles
ratios.

\begin{figure}[h]
  \includegraphics[width=5.9cm]{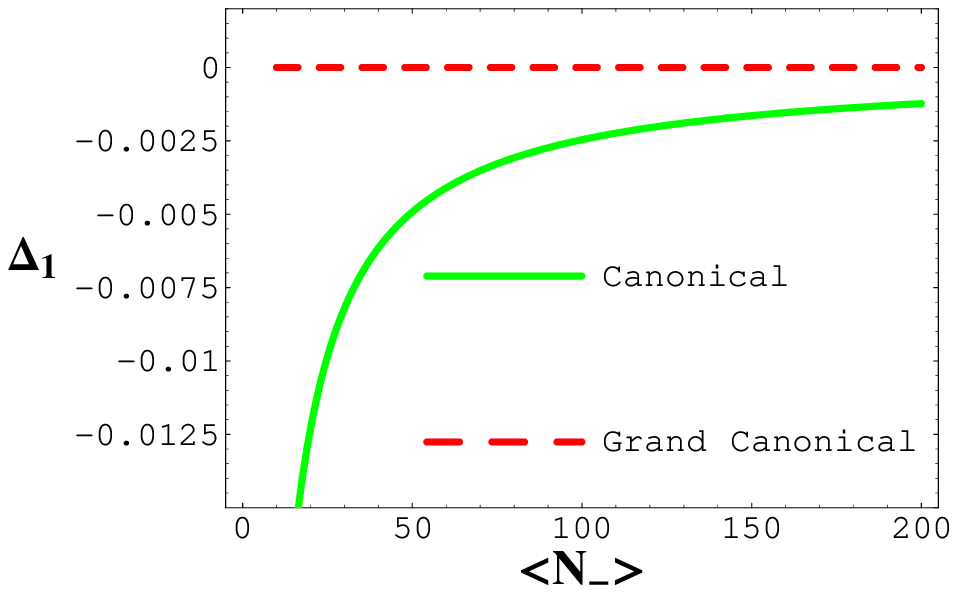}
  \includegraphics[width=5.9cm]{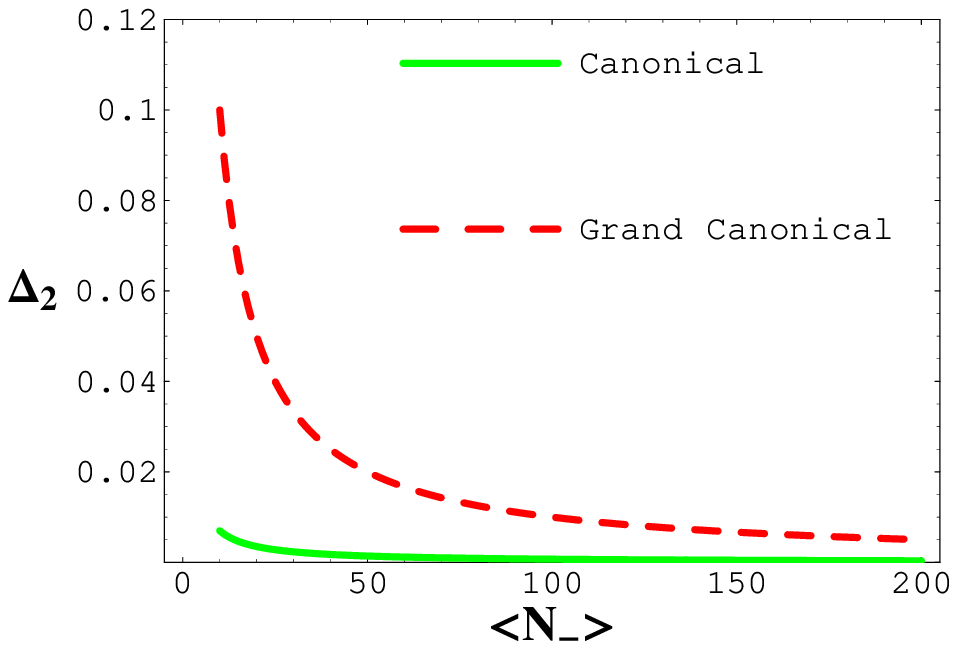}
  \includegraphics[width=5.9cm]{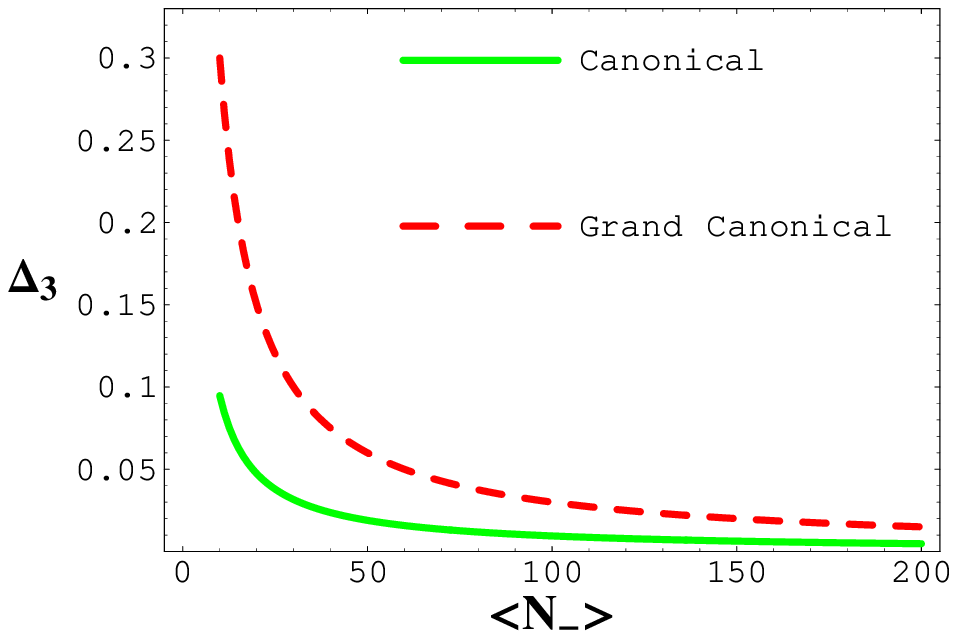}
  \caption{Deviations of the first $\Delta_1$, second $\Delta_2$
  and the third $\Delta_3$ charged particle moments from the thermodynamic limit as
  functions of charge particle multiplicities $\langle N_-\rangle$ calculated at fixed $\langle
N_-\rangle/\langle N_+\rangle=0.8$. The results are shown for the canonical and the
grand canonical ensembles.} \label{moments}
\end{figure}

In heavy ion collisions at RHIC charged particle ratios like e.g. $\pi^-/\pi^+$,
$K^-/K^+$ or $\bar{p}/p$ has been measured (see e.g. \cite{ratios}) in the range
between 0.6 to 1.0. Thus, for RHIC conditions one estimates  from Eq. (\ref{var
ratio}) that the relevant domain of $q/z_0$   is between 0 and 0.5. In
Fig.~\ref{moments} we calculate from Eqs. (\ref{new1}) and (\ref{new2}) the NLO
corrections to the first, the second and the third charge particle density moments
for canonical and grand canonical ensemble. These corrections are plotted as a
function of the multiplicity of negatively charged particles $\langle N_-\rangle$ at
fixed value of $\langle N_-\rangle/\langle N_+\rangle =0.8$ particles ratio.

It is specific for the grand canonical distribution that the first moment $\Delta_1$
coincides with its thermodynamic limit. However, as seen in Fig.~\ref{moments} this
is not anymore the case for higher moments where deviations from thermodynamic limit
in the GC ensemble are not negligible even for $\langle N_-\rangle\simeq 150$. It is
also clear from Fig. \ref{moments} that for fixed $\langle N_-\rangle$ the
difference between GC and C ensemble increases with increasing order of particle
moments. The convergence of $\Delta_k$ to their thermodynamic limit values  is
slower with increasing $k$, for both ensembles. Finally, it is clear from
Fig.~\ref{moments} that the first moment is converging from below whereas the second
and the third from the above to their asymptotic values.

 With the above results for the NLO corrections to different particle density
moments one can also study the convergence and the thermodynamic limit of the scaled
variance \eqref{eaq30}.

Indeed from Eqs. \eqref{eaq31} and \eqref{eaq32} one finds:

\begin{equation}\label{eaq39}
    \text{T-}\lim \omega_N=
    \frac{\mathcal{R}_{2;q}-2\mathcal{R}_{1;q}\langle n\rangle_{\infty}}{\langle
    n\rangle_{\infty}}\,.
\end{equation}
Applying in  Eq. \eqref{eaq39}  the C and GC correction factors $\mathcal{R}_{k;q}$
from Eqs. \eqref{eaq33} and \eqref{eaq34} one gets in the canonical ensemble

\begin{subequations}
  \begin{equation}
    \text{T-}\lim\mathcal{\omega}_N^{C}=
    \frac{1}{2}\frac{\sqrt{q^2+4z_0^2}\mp q}{\sqrt{q^2+4z_0^2}}\,,
\end{equation}
while in the grand canonical system
\begin{equation}
    \text{T-}\lim\mathcal{\omega}_N^{GC}
   =1\,.
\end{equation}
\end{subequations}
Thus, the scaled variance is finite in the T-limit although it differs in the C and the GC
ensembles. This result agrees with the previous finding in \cite{begun} and explains how
to reconcile this with the T-limit equivalence of different statistical ensembles.

The scaled variance is not the only example  of  semi--intensive
variable. There is  actually  a broad  class of variables

\begin{subequations}
  \begin{equation}
    \mathcal{S}_k=\frac{\langle N^k\rangle - \langle N\rangle^k}{\langle N\rangle^{k-1}}
\end{equation}
which have the same properties as $\omega_N$:  they are finite in T-limit and have
different values dependently on how the charge conservation is implemented in the
description of the  system. Indeed from \eqref{eaq32} one gets
\begin{equation}
    \text{T-}\lim\mathcal{S}_k=
    \frac{\mathcal{R}_{k;q}-k\mathcal{R}_{1;q}\langle n\rangle_{\infty}^{k-1}}{\langle
    n\rangle_{\infty}^{k-1}}\,,
\end{equation}
\end{subequations}
thus,  following   Eqs. (\ref{eaq33}--\ref{eaq34}) the  C and GC values for
positive(negative) particles in the T-limit are found as
\begin{subequations}
  \begin{equation}
    \text{T-}\lim\mathcal{S}_k^{C}=
    \frac{k(k-1)}{4}\frac{\sqrt{q^2+4z_0^2}\mp q}{\sqrt{q^2+4z_0^2}}\,,
\end{equation}
while in the grand canonical ensemble
\begin{equation}
    \text{T-}\lim\mathcal{S}_k^{GC}
   =\frac{k(k-1)}{2}\,.
\end{equation}
\end{subequations}
The scaled variance $\omega_N$ is just a  special case of $S_k$ corresponding to  $k=2$.

Another example is  a broad class of variables closely related to cumulant or factorial
cumulant moments $\kappa_p$ defined as \cite{thi}

\begin{subequations}
  \begin{eqnarray}
    \ln{\mathcal{G}(\e^\lambda,V)}&=&\sum\limits_{p}\kappa_p\frac{\lambda^p}{p!}\qquad\text{for cumulant
    moments}\,,\\
\ln{\mathcal{G}(\lambda,V)}&=&\sum\limits_{p}\kappa_p\frac{\lambda^p}{p!}\qquad\text{for
factorial cumulant moments}\,.
\end{eqnarray}
\end{subequations}
In the T-limit those moments  are linear in $V$, thus the ratios
\begin{equation}
    \mathcal{K}_{p;r}=\frac{\kappa_p}{\kappa_r}\,,
\end{equation}
are obviously semi--intensive quantities. For cumulant moments the quantity
$\mathcal{K}_{2;1}$ coincides with the scaled variance \eqref{sc var} while for factorial
cumulant moments it is equal to $\omega_N-1$.

One can also construct more involved semi-inclusive variables having a finite T-limit
behavior which are determined by higher order asymptotic terms of the corresponding
probability distributions.
\section{Summary and conclusions}
 We have considered   statistical systems with global  charge conservation and
discussed   the properties of different charged particle probability distributions. We
have put particular emphasis on the limiting behavior of the probability densities in
their approach towards  the thermodynamic limit. In particular we have calculated the
asymptotic values and the first subleading finite volume corrections.  Our results were
obtained in the statistical model of quasi non--interacting gas of charged particles,
constrained by the conservation laws. Such a model has been recently shown to be
successful in  describing   particle production in heavy ion collisions and lattice QCD
thermodynamics restricted to the confined, hadronic phase.

 We have discussed  the differences in the asymptotic properties of the probability
functions for a system with an exact, that is canonical, (C) and with an average, that is
grand canonical, (GC) implementation of  charge conservation. We have shown that in the
thermodynamic limit the corresponding probability distributions in the GC and C ensembles
coincide and are described as  generalized functions. This property is a direct
consequence of the GC and C ensemble equivalence in the thermodynamic limit. However, the
first finite volume corrections to the asymptotic value  differ  for both ensembles.

Finally, using the results of the probability functions we have derived the
asymptotic behavior  of the charged particle moments and established the differences
in the  GC and C formulation. We have also applied these results to find the
thermodynamic limit of a class of  semi--intensive quantities. It was shown that in
systems with exact and average charge conservation such  quantities should naturally
converge to different values in the thermodynamic limit. This is because the
behavior  of the semi--intensive quantities in the near vicinity to the
thermodynamic limit are determined by the subleading, finite volume, corrections to
the probability distributions which are specific to a given statistical ensemble.

Similar conclusions related to the thermodynamic limit in different ensembles can be
also drawn for nonrelativistic systems \cite{prepar}. Such an analysis could be of
interest  for low energy processes in which the  particle number conservation is
preserved.

\begin{acknowledgments}
We  acknowledge the stimulating discussions with M.~Gazdzicki. This work is partially
supported by the Polish Committee for Scientific Research under contract
KBN~2~P03B~069~25 and the Polish--South African Science and Technology cooperation
project.
\end{acknowledgments}

\appendix
\section{Leading and NLO coefficients of $\langle N^k\rangle^C_Q$}
\label{C asympt}%
The leading and next to leading order contributions  to the particle moments $\langle
N^k\rangle^C$  are obtained as the coefficients $c_k$ and $c_{k-1}$ at large volume
expansion terms in Eq. \eqref{gen fn mom}, i.e. from
\begin{equation}\label{eqA1}
  \langle N^k\rangle^C = c_k\,V^k+c_{k-1}\,V^{k-1}+\mathcal{O}(V^{k-2})\,.
\end{equation}
To get these coefficients using the expansion for the Bessel functions given in
\eqref{bess as} we first observe that the second and the third terms in the curly bracket
in Eq. \eqref{gen fn as}  can be neglected. This is because the contributions of these
terms to the $O(V^{k-1})$ order coefficient exactly cancel each other since the
derivative in Eq. \eqref{gen fn mom} is taken at $\lambda =1$. Thus, it is sufficient to
consider
\begin{equation}\label{eqA2}
   \mathcal{G}(\lambda,V)\simeq f(\lambda)e^{Vg(\lambda )}
\end{equation}
with
\begin{equation*}
    f(\lambda )=\sqrt{\frac{x_q}{x_{q;\lambda}}}\,, \qquad
    g(\lambda)= x_{q;\lambda} -x_q+q\ln\frac{q+x_q}{q+x_{q;\lambda}}\,.
\end{equation*}
where $x_q$ is as in Eq. \eqref{coeff x}.

To calculate  $c_k$ and $c_{k-1}$   we use the identity
\begin{equation}\label{eqA3}
    \left(\lambda\frac{d}{d\lambda}\right)^k  \mathcal{G}   =\lambda^k\frac{d^k}{d\lambda^k} \mathcal{G} +
\frac{k(k-1)}{2}\lambda^{k-1}\frac{d^{k-1}}{d\lambda^{k-1}}
\mathcal{G} + \cdots
\end{equation}
and keep  the first two terms with the highest order derivatives.
Only the first term in Eq. \eqref{eqA3} contributes to $c_k$

\begin{equation}\label{eqA4}
c_k=f(\lambda)\left(\frac{d g}{d\lambda}\right)^k
\e^{Vg(\lambda)}\,.
\end{equation}
However, the $c_{k-1}$ coefficient  is receiving contributions from both terms in Eq.
\eqref{eqA3}.  The highest derivative term in Eq. \eqref{eqA3} is calculated from

\begin{equation}\label{eqA5}
f(\lambda)\frac{d^k}{d\lambda^k}\e^{Vg(\lambda)}+ {k\choose{k-1}}\frac{d f}{d\lambda}
\frac{d^{k-1}}{d\lambda^{k-1}}\e^{Vg(\lambda)} +  \cdots \,\,,
\end{equation}
where  up to $\mathcal{O}(V^{k-3})$
\begin{equation}\label{eqA6}
\begin{split}
   & \frac{d^k}{d\lambda^k}\e^{Vg(\lambda)}\simeq
V^k \left(\frac{dg}{d\lambda}\right)^k
\e^{Vg(\lambda)}+V^{k-1}d_k\e^{Vg(\lambda)}\simeq
\frac{d}{d\lambda}\left[\frac{d^{k-1}}{d\lambda^{k-1}}\e^{Vg(\lambda)}\right]\simeq\\
&\
\frac{d}{d\lambda}\left[V^{k-1}\left(\frac{dg}{d\lambda}\right)^{k-1}
\e^{Vg(\lambda)}+V^{k-2}d_{k-1}\e^{Vg(\lambda)}\right]\,,
\end{split}
\end{equation}
The coefficients $d_k$ obey  the  following recursion
relation

\begin{equation}\label{eqA7}
d_k=d_{k-1}\frac{d g}{d\lambda}+(k-1)\left(\frac{d
g}{d\lambda}\right)^{k-2}\frac{d^2g}{d\lambda^2}\,,
\end{equation}
 with the
solution
\begin{equation}\label{eqA8}
d_k=\frac{k(k-1)}{2}\left(\frac{dg}{d\lambda}\right)^{k-2}\frac{d^2
g}{d\lambda^2}\,.
\end{equation}
 Applying the above result for $d_k$ coefficients in Eq. \eqref{eqA6} and using  Eq.
\eqref{eqA3} one gets the final expression for the $c_{k-1}$ coefficient

\begin{equation}\label{eqA9}
 c_{k-1}=   \frac{k(k-1)}{2}f(\lambda)\left(\frac{d g}{d\lambda}\right)^{k-1} +
    k\frac{d f}{d\lambda}\left(\frac{d g}{d\lambda}\right)^{k-1} +
\frac{k(k-1)}{2}f(\lambda)\left(\frac{dg}{d\lambda}\right)^{k-2}\frac{d^2
g}{d\lambda^2}\,,
\end{equation}
where the last term is obtained from the $(k-1)$--order derivative in Eq. \eqref{eqA3}.

\section{Grand Canonical charge density probability distribution in the T-limit}
Let us consider the grand canonical  probability distribution to
find  the charge density $ q$  in a system of volume $V$ and
average charge density  $\langle  q\rangle$. This probability,
following Eq. \eqref{prob GC dens}, reads
\begin{equation}\label{eqB1}
  {\mathbf{P}}^{GC}_{\langle  q\rangle}( q,V)  =
    V\e^{-Vx_{\langle q\rangle}}
    I_{V q}(2V z_0)\left[\frac{\langle  q\rangle +
    x_{\langle q\rangle}}{2 z_0}\right]^{V q}\,.
\end{equation}
with $x_{\langle q\rangle}$ as  in Eq. \eqref{coeff x} but with
$q$ being replaced by $\langle q\rangle$.

Our goal is to find  the first subleading contribution to $ {\mathbf{P}}^{GC}_{\langle
q\rangle}( q,V) $ in
 the thermodynamic limit.

It is rather straightforward to see from Eq. \eqref{eqB1} that the limit $V\to\infty$ for
fixed $q$ does not exist as a regular function \cite{lt}. However, this limit can be
found as a generalized function. To show it, we consider  the integral
\begin{equation}
   \langle G\rangle = \int d q G( q){\mathbf{P}}^{GC}_{\langle
    q\rangle}(q,V)\,.\label{eqB2}
\end{equation}
where the probability distribution ${\mathbf{P}}^{GC}_{\langle q\rangle}$ is smeared out
with a test function $G$.

In our previous study   \cite{lt} it was shown that in the T-limit
\begin{equation}\label{eqB3}
    \langle G\rangle = G(\langle  q\rangle) +
    \mathcal{O}(V^{-1})\,.
\end{equation}
Thus, one can write
\begin{equation}\label{eqB4}
    \lim_{V\to\infty}{\mathbf{P}}^{GC}_{\langle  q\rangle}( q,V) =
    \delta\left( q-\langle q\rangle\right)\,.
\end{equation}
We  apply the asymptotic form of the $I_\alpha$ Bessel function
\eqref{bess as} in \eqref{eqB1} and then calculate the integral
\eqref{eqB2} using the saddle-point method. To obtain the NLO term
of this integral one should take into account contributions coming
from the probability \eqref{eqB1} in the next to leading order and
corresponding contributions from \eqref{eqC3} simultaneously.

From \eqref{eqB1} and (\ref{bess as}, \ref{coeff x}) one gets
\begin{equation}\label{eqB5}
\begin{split}
      {\mathbf{P}}^{GC}_{\langle  q\rangle}( q,V)\simeq
      & V^{1/2}\frac{e^{-V(x_{\langle q\rangle}-x_q)}}
    {\sqrt{2\pi x_q}}\left[\frac{\langle q\rangle+x_{\langle q\rangle}}
    { q+x_q}\right]^{V q} +\\
   & V^{-1/2}\frac{6 z_0^2 -  q^2}
    {12\sqrt{2\pi}x_q^{7/2}}\left[\frac{\langle q\rangle+x_{\langle q\rangle}}
    { q+x_q}\right]^{V q}
    \e^{-V\left(x_{\langle q\rangle}-x_q\right)}\,.
\end{split}
\end{equation}

The last term in Eq. \eqref{eqB5} contributes to the order $\mathcal{O}(V^{-1})$  in $
{\mathbf{P}}^{GC}_{\langle q\rangle}( q,V)$ as
\begin{equation}\label{eqB6 }
    \frac{1}{V}\frac{6 z_0^2 -  q^2}
    {12\sqrt{2\pi}x_q^3}\delta( q - \langle q\rangle)\,.
\end{equation}
A further contribution of   order $\mathcal{O}(V^{-1})$ comes from the coefficient $c_1$
calculated from  Eq. \eqref{eqC3} as
\begin{equation}
\label{eqB7 } c_1
=\frac{\sqrt{\pi}}{4}\left(\frac{d}{dy}\right)^{2}
\left.\left[\frac{G(y)}
{\sqrt{2\pi}(y^2+4z_0^2)^{1/4}}\left(\frac{S(y_0)-S(y)}{(y-y_0)^2}\right)^{-3/2}\right]\right\vert_{y=y_0}\,.
\end{equation}
with
\begin{equation}\label{eqB8 }
      S(q)=x_q-x_{\langle q\rangle}  +
 q\ln\frac{\langle q\rangle+x_{\langle q\rangle}}{q + x_q}\,.
\end{equation}
This gives
\begin{equation}\label{eqB9 }
    c_1 = \frac{1}{12}\frac{\langle q \rangle^2 - 6 z_0^2}{x_{\langle q\rangle}^{3}} G(\langle
     q\rangle) +
    \frac{x_{\langle q\rangle}}{2}G^{''}(\langle q\rangle)\,.
\end{equation}
Consequently, the  leading and the subleading contributions to ${\mathbf{P}}^{GC}_{\langle
q\rangle}(q,V)$ are obtained as
\begin{equation}\label{eqB10 }
{\mathbf{P}}^{GC}_{\langle  q\rangle}(q,V)=
  \delta\left( q-\langle q\rangle\right) +
  \frac{x_{\langle q\rangle}}{2 V }\,\delta^{''}( q -\langle q\rangle)
  +\mathcal{O}(V^{-2})\,,
\end{equation}
where the double prime in the delta function is the second derivative with respect to the
charge density $q$.

\section{Watson-Laplace theorem}\label{saddle}
Let us consider the  Laplace integral
\begin{equation}\label{laplace1}
    F(\lambda)=\int\limits_I dx \e^{\lambda S(x)}f(x)\,.
\end{equation}
An asymptotic expansion of   $F(\lambda)$ in the limit $\lambda\to
\infty$ is given by the classical Watson-Laplace theorem:
\begin{quotation}
  Let $I=[a,b]$ be a finite interval such that
  \begin{enumerate}
    \item $\max\limits_{x\in I} S(x)$ is reached only in the single point $x_0=a,\ a<x_0<b$.
    \item $f(x),S(x)\in C(I)$.
    \item $f(x), S(x)\in C^\infty$ in the vicinity of $x_0$, and $S^{''}(x_0)\neq
    0$\,.
  \end{enumerate}
  Then, for  $\lambda\in S_\epsilon$ and  $\lambda\to\infty$,\,  the following expansion holds
  \begin{equation}\label{laplace}
    F[\lambda]\simeq \e^{\lambda S(x_0)}\sum\limits_{k=0}^\infty c_k\lambda^{-k-1/2}\,,
 \end{equation}
 where the coefficients
\begin{equation}\label{eqC3}
    c_k =\frac{\Gamma(k+1/2)}{(2k)!}\left(\frac{d}{dx}\right)^{2k}
\left.\left[f(x)\left(\frac{S(x_0)-S(x)}{(x-x_0)^2}\right)^{-k-1/2}\right]\right\vert_{x=x_0}\,.
\end{equation}
 $S_\epsilon$~is here a segment $|\arg z|\leqslant\frac{\pi}{2}-\epsilon<\frac{\pi}{2}$ in
the complex $z$-plane.
\end{quotation}

\end{document}